\def \SAIT #1 #2 {{\em Mem.\ Soc.\ Astron.\ It.\/} {\bf #1}, #2}
\def \MESS #1 #2 {{\em The Messenger\/} {\bf #1}, #2}
\def \ASTRNACH #1 #2 {{\em Astron. Nach.\/} {\bf #1}, #2}
\def \AAP #1 #2 {{\em Astron. Astrophys.\/} {\bf #1}, #2}
\def \AAL #1 #2 {{\em Astron. Astrophys. Lett.\/} {\bf #1}, L#2}
\def \AAR #1 #2 {{\em Astron. Astrophys. Rev.\/} {\bf #1}, #2}
\def \AAS #1 #2 {{\em Astron. Astrophys. Suppl. Ser.\/} {\bf #1}, #2}
\def \AJ #1 #2 {{\em Astron. J.\/} {\bf #1}, #2}
\def \ANNREV #1 #2 {{\em Ann. Rev. Astron. Astrophys.\/} {\bf #1}, #2}
\def \APJ #1 #2 {{\em Astrophys. J.\/} {\bf #1}, #2}
\def \APJL #1 #2 {{\em Astrophys. J. Lett.\/} {\bf #1}, L#2}
\def \APJS #1 #2 {{\em Astrophys. J. Suppl.\/} {\bf #1}, #2}
\def \APSS #1 #2 {{\em Astrophys. Space Sci.\/} {\bf #1}, #2}
\def \ASR #1 #2 {{\em Adv. Space Res.\/} {\bf #1}, #2}
\def \BAIC #1 #2 {{\em Bull. Astron. Inst. Czechosl.\/} {\bf #1}, #2}
\def \JSQRT #1 #2 {{\em J. Quant. Spectrosc. Radiat. Transfer\/} {\bf #1}, #2}
\def \MN #1 #2 {{\em Mon. Not. R. Astr. Soc.\/} {\bf #1}, #2}
\def \MEM #1 #2 {{\em Mem. R. Astr. Soc.\/} {\bf #1}, #2}
\def \PLR #1 #2 {{\em Phys. Lett. Rev.\/} {\bf #1}, #2}
\def \PASJ #1 #2 {{\em Publ. Astron. Soc. Japan\/} {\bf #1}, #2}
\def \PASP #1 #2 {{\em Publ. Astr. Soc. Pacific\/} {\bf #1}, #2}
\def \NAT #1 #2 {{\em Nature\/} {\bf #1}, #2}

\documentstyle{memsait}
\input epsf.sty
\begin{opening}
\title{RAPID VARIABILITY OF BLAZARS} 

\author{MARCO CHIABERGE$^1$, GABRIELE GHISELLINI$^2$}
\institute{$^1$ S.I.S.S.A., via Beirut 2-4, 34014 Trieste, Italy\\
$^2$ Osservatorio Astronomico di Brera, Via Bianchi 46, 22055 Merate, Italy}
\date{} 
\end{opening}

\begin{document}

\oddpagefooter{}{}{} 
\evenpagefooter{}{}{} 
\ 
\bigskip

\begin{abstract}
Blazars are characterized by large amplitude and fast variability, indicating that the 
electron distribution is rapidly changing, often on time scales shorter than the light 
crossing time.  We study the time dependent behavior of the electron distribution after
episodic electron injection phases, and calculate the observed synchrotron and self 
Compton radiation spectra.  Since photons produced in different part of the source have 
different travel times, the observed spectrum is produced by the electron distribution 
at different stages of evolution.  
Time delays between the light curves of fluxes at different 
frequencies are possible, as illustrated for the specific case of the BL Lac object 
Mkn 421.
\end{abstract}

\section{Introduction}

Variability is one of the defining properties of blazars, characterized  by 
variations of their flux even of two orders of magnitude in time scales of 
years, and smaller changes, but still up to factor 2, in hours/days time
scales.
Light curves as seen, e.g., in the X--rays and in the optical, 
often show a quasi--symmetric behavior, with rise and decay time scales approximately 
equal (see i.e. Urry et al. 1997, Ghisellini et al. 1997, Massaro et al. 1996,
Giommi et al. 1998), indicating that both times are connected to
the light travel time across the source $R/c$, and therefore suggesting that
the cooling times of the emitting electrons are shorter. Furthermore this implies that 
the emitting electron distribution is significantly changing on time scales shorter than 
$R/c$, at least at these energies. In order to reproduce the observed variability
pattern we need to study the time evolution of the emitting particles distribution 
{\it and} to take into account the different light travel times of photons produced in 
different regions of the source. 
The observed flux is then, at any time, {\it the sum of the emission produced by
particle distributions of different ages}, each one produced in a different region
of the source: even a homogeneous source then resembles an inhomogeneous one.  
In \S2 we shortly present the model assumptions, in \S3 we show some results
of the simulations in the case of electron distributions homogeneously injected 
throughout the source and in a more realistic case of a shock traveling down a region
of a jet. In \S4 we apply our model to the specific case of MKN 421.

\section{The model}

We assume that the emission is produced by a distribution
of relativistic electrons injected in a region of typical dimension
$R$ embedded in a tangled magnetic field $B$, at a rate $Q(\gamma)$ 
[cm$^{-3}$ s$^{-1}$] ($\gamma$ is the electrons Lorentz factor). 
Electrons lose energy by emitting synchrotron and synchrotron self-Compton
radiation (SSC). Escape from the source is also considered ($t_{esc}$ is assumed being
independent of energy). 
The continuity equation governing the temporal evolution of the electron distribution
$N(\gamma,t)$ [cm$^{-3}$] is 
\begin{equation}
\frac{\partial N(\gamma,t)}{\partial t} = \frac{\partial}{\partial\gamma} 
\left[ \dot\gamma(\gamma,t) N(\gamma,t)\right] + Q(\gamma,t) - 
\frac{N(\gamma,t)}{t_{esc}}
\label{cont}
\end{equation}
where $\dot\gamma= \dot\gamma_{S} + \dot\gamma_{C}$ is the total cooling rate (synchrotron
and synchrotron self--Compton).
We calculate SSC spectra produced by the calculated electrons distribution at any 
given time. 
The details of the numerical method used to solve the equation are illustrated
in Chiaberge \& Ghisellini (1998). 
If the particle distribution evolves more rapidly than
$R/c$ {\it the observer will see, at any time, a convolution of different
spectra, each produced in a different part of the source}. 
Initially the observer only sees the emission coming from fresh electrons
located in the region (`slice') closest to her/him; then the inner parts
of the source become visible, also showing `young' spectra, while
electrons in the front slices are evolving.
After a time $R/c$ all the emitting region will be visible:
the back of it with fresh electrons and the front of it with older electrons.
In order to take into account this effect, we divide the source of size $R$ into 
$n$ slices of equal thickness $R_{sl} < t_{min}$ ($t_{min}$
is the shorter among cooling, injection and escape times). In this way, each single slice 
can be considered as an homogeneous emitting region.
We then sum the contribution of each slice at any time.
We consider slices having equal volumes
(assuming a `cubic' geometry), with the line of sight placed at 90$^\circ$
with respect to one face of the cube.
This angle, in the lab frame, transforms to a viewing angle of $\sim 1/\Gamma$,
if the source moves with a Lorentz factor $\Gamma$, which is
appropriate for blazars.
Extension to different geometries (i.e. cylinder, sphere, etc)
is trivial, by properly weighting each slice volume and does not affect the results.
Even in the simplest `cubic' case time-lags among 
light curves at different frequencies are observable.



%

\section{Discussion}

\begin{figure}
\epsfysize=13cm 
\hspace{3.5cm}\epsfbox{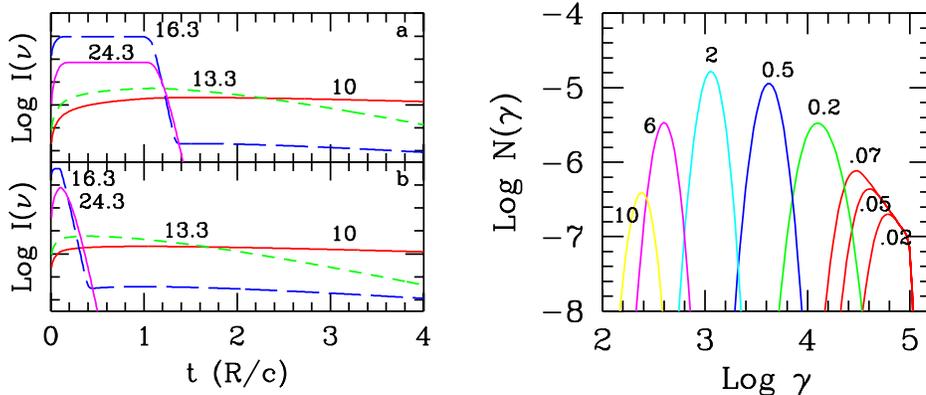} 
\vspace{-7.7cm}
\caption[h]{{\it Left:} light curves of the specific intensity at different frequencies, 
as observed in the comoving frame, in the case of Gaussian injection and 
$t_{inj}\ll R/c$, to illustrate
light crossing time effects, included in a) and ignored in b).
Labels correspond to the logarithm of the frequency. 
For clarity, the intensity of each light curve has been multiplied by 
different constants. {\it Right:} time evolution of the particle distribution $N(\gamma)$
corresponding to an injection of particles distributed in energy
as a Gaussian, centered at $\gamma=10^5$, for $t_{inj}=0.1 R/c$.
Labels indicate time after the beginning of the injection, 
in units of $R/c$.}
\label{fig1}
\end{figure}
\begin{figure}
\epsfysize=13cm 
\hspace{3.5cm}\epsfbox{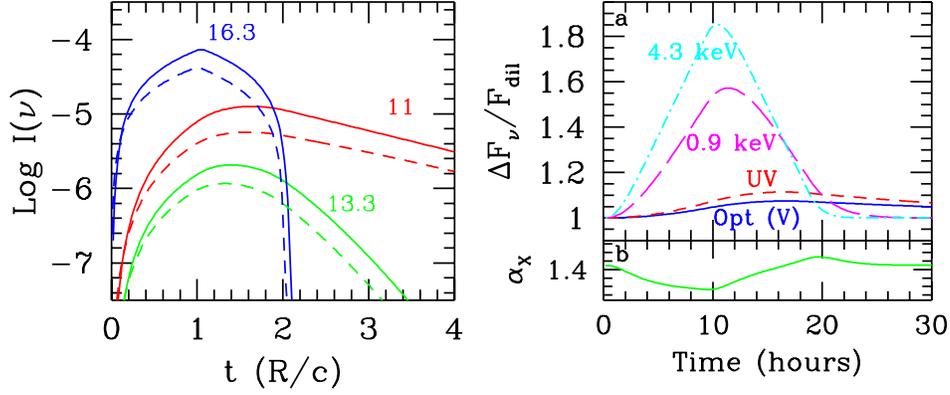} 
\vspace{-7.5cm}
\caption[h]{{\it Left:} simulated light curves at different frequencies in the case of 
power law injection lasting for a time $t_{inj}=R/c$ (dashed lines, homogeneous 
injection case) and  in the case of a shock active for a time $t=R/c$ traveling across a 
region of dimension $R$ (solid lines). {\it Right:} simulated light curves at different 
frequencies (a) and X--ray spectral index (b) for the Mkn 421 May 1994 flare.}
\label{fig2}
\end{figure}

We summarize here the results of the simulations of two different illustrative 
injection phases:
i) narrow Gaussian electron distribution centered at $\gamma=10^{5}$ lasting
for a time $t_{inj} \ll R/c$;
ii) power law electron distribution ($Q(\gamma) \propto \gamma^{-p}$, $p=1.7$) 
between $\gamma_{min}=1$ and $\gamma_{max}=10^5$, lasting for a time $t_{inj}=R/c$.
The source physical parameters are: $R=10^{16}$ cm,
$B=1$ Gauss, injected luminosity $L_{inj}=4 \times 10^{41}$ erg s$^{-1}$, $t_{esc}=1.5 R/c$.
We stress that our main purpose is to reproduce the quasi--symmetric light curves 
often observed at optical--X rays frequencies, in order to put strict constraints 
on the physical parameters of a source producing such a variability pattern. 
The main difference between injecting a Gaussian distribution of electrons
and injecting a power--law (with $p>0$) is that in the first case the emission will be 
concentrated first at high frequencies, and only after some $t_{cool}$ 
electrons can substantially emit at lower frequencies. This produces a time delay 
between the peaks of the emission at different frequencies in case (i) (see fig. \ref{fig1}). 
Also notice the different behavior of the light curves at different frequencies, 
showing plateau at the highest corresponding electron energies.
In case (ii) at frequencies where $t_{cool} \ll R/c$ both the rise and the decay are 
controlled by the light crossing time, and we have symmetric light curves with
time lags depending on the different cooling times.\\
According to our results, quasi symmetric light curves  can be only originated in 
two cases: 1) $t_{inj}\ll R/c \sim t_{cool}$;
2) $t_{inj}\sim R/c$ and $t_{cool}\ll R/c$.
In case (1) symmetric light curves are present only within a very small
range of frequencies ($t_{cool}\sim R/c$), while in case (2) quasi--symmetric 
light curves can occur at all frequencies corresponding to particle cooling 
time scales shorter than $R/c$. 
Furthermore the second case can be interpreted 
as a result of a shock lasting for a time $t_{inj}\sim R/c$. In fact, we reproduced  
the case of a shock of longitudinal dimension $R$ and width $r_s \ll R$
running along a region of the jet of same dimension $R$ (perpendicular to the 
jet axis). As in the previous cases, we assume that the observer is located 
at an angle $1/\Gamma=\sqrt{1-\beta^2}$ from the jet axis, such that 
the viewing angle in the comoving frame is $90^\circ$. We calculate the observed 
light curves in the comoving frame, by summing the contributions of each part of the source, 
taking into account  both the traveling of the shock across the source 
(accelerating particles in different
regions at different times, as the shock travels across the source) and the light 
travel time effect. Details are presented in Chiaberge \& Ghisellini (1998). 
The light curves obtained in this shock case are similar to the corresponding 
`homogeneous' case ($t_{inj}=R/c$), with slightly longer time--lags (see fig. \ref{fig2})

\section{Application to MKN 421}

In May 1994 the ASCA satellite revealed an X--ray flare of the nearby ($z=0.03$) 
BL Lac object (Takahashi et al. 1996) during an high state of TeV emission 
(Macomb et al. 1995). 
Observations report an increase of a factor $\sim 2$ of the 2--10 keV flux,
with a doubling time scale of $\sim 12$ hours. Much less amplitude
variability is present in the IR, optical, UV and GeV bands. 
Takahashi et al. (1996) found a time-lag between hard X--rays 
($2-7.5$ keV) and soft X--rays ($0.5-1.5$ keV) of $\sim 1$ hour: the hard 
X--rays lead the soft X--rays.
They interpret this as due to synchrotron cooling. We can qualitatively reproduce this 
behavior, assuming that the rapid variability is due to the sum of a rapidly evolving 
component and a quasi-constant one, corresponding to the high state fit of the spectral
energy distribution (Chiaberge \& Ghisellini, 1998). 
We take into account the effects of beaming using the following transformations: 
if $\Gamma$ is the bulk Lorentz factor, $\theta$ the viewing angle and
$\delta=[\Gamma(1-\beta \cos\theta)]^{-1}$ the beaming factor,
the observed intensity is $I(\nu)=\delta^3 I^\prime (\nu^\prime)$
and $t=t^\prime/\delta$,
where $I^\prime (\nu^\prime)$ and $t^\prime$ are the comoving intensity
and comoving time scales, respectively.
We found the following parameters for the variable component:
$R=1.5\times 10^{16}$ cm, $B=0.13$ Gauss, $\delta=15.5$, 
$\ell_{inj}=1.5 \times 10^{-3}$, $Q(\gamma)\propto \gamma^{1.4} 
\exp(-\gamma/\gamma_{max})$ between 
$\gamma_{min}=10^3$ and $\gamma_{max}=8.5\times 10^5$. We perform the
simulation in the shock case with 
$r_s= 0.1 R/c$ (width of the shock), $t_{s}= R/c$ (time during which the shock is active)
and $\beta^{\prime}_s \sim 1$ (velocity of the shock in the comoving frame). Light curves 
and X--ray spectral index variability are reported in fig \ref{fig2}.

\end{document}